\documentclass[aps,prd,twocolumn,superscriptaddress,nofootinbib,floatfix]{revtex4-2}
\usepackage{amsmath,amssymb,bm}
\usepackage{graphicx}
\usepackage{booktabs}
\usepackage{xcolor}
\usepackage{enumitem}
\usepackage[colorlinks=true,linkcolor=blue!60!black,citecolor=blue!60!black,urlcolor=blue!60!black]{hyperref}

\newcommand{\pd}{\partial}
\newcommand{\dd}{\mathrm{d}}
\newcommand{\eps}{\epsilon}
\newcommand{\Dphi}{\Delta_\phi}
\newcommand{\ord}[1]{\mathcal{O}(\partial^{#1})}

\newcommand{\ETAOTWOFIVE}{-0.1324(27)}
\newcommand{\ETAOTWOFOUR}{-0.3244(38)}
\newcommand{\ETAOFOURFIVETOT}{-0.153(7)}
\newcommand{\ETAOFOURTHREETOT}{-0.575(9)}
\newcommand{\ETAOTWOTHREE}{-0.562(12)}
\newcommand{\SIGMATHREE}{0.076}
\newcommand{\NUCTHREE}{0.359}
\newcommand{\ETAOFOURFOURTOT}{-0.352(9)}

\newcommand{\APMSOTWOFIVE}{2.7}
\newcommand{\EPMSOTWOFIVE}{-0.1328}
\newcommand{\APMSOTWOFOUR}{1.9}
\newcommand{\EPMSOTWOFOUR}{-0.3245}
\newcommand{\APMSOTWOTHREE}{1.5}
\newcommand{\EPMSOTWOTHREE}{-0.5655}

\newcommand{\EPMSOFOURFOUR}{-0.3514}

\newcommand{\DPMSMAXOTWO}{0.004}

\newcommand{\DSIGMAOTWO}{2.650(13)}
\newcommand{\DSIGMAOFOUR}{2.653(10)}

\newcommand{\DSIGMA}{2.65(1)}

\newcommand{\ETADSIGMA}{-0.65}

\begin{document}

\title{Yang--Lee edge singularity at $\mathcal{O}(\partial^4)$ from the functional renormalization group: revised analysis
}

\author{Konrad Kockler}
\affiliation{Institut f\"ur Theoretische Physik, Universit\"at Heidelberg, Philosophenweg 16, 69120
Heidelberg, Germany}

\author{Jan M.~Pawlowski}
\affiliation{Institut f\"ur Theoretische Physik, Universit\"at Heidelberg, Philosophenweg 16, 69120
Heidelberg, Germany}
\affiliation{ExtreMe Matter Institute EMMI, GSI Helmholtzzentrum für Schwerionenforschung, Planckstrasse 1, 64291 Darmstadt, Germany}

\author{Vladimir V. Skokov}
\affiliation{Department of Physics, North Carolina State University, Raleigh, NC 27695, USA}
\affiliation{ExtreMe Matter Institute EMMI, GSI Helmholtzzentrum für Schwerionenforschung, Planckstrasse 1, 64291 Darmstadt, Germany}

\date{\today}

\begin{abstract}
  We determine the critical exponents of the Yang--Lee edge singularity as continuous functions of dimension using the functional renormalization group, retaining all operators at second and fourth order in the derivative expansion. At fourth order, we obtain $\eta=\ETAOFOURFIVETOT$, $\ETAOFOURFOURTOT$ and $\ETAOFOURTHREETOT$ in $d=5$, $4$ and $3$, respectively, in agreement with the six-loop $\eps$ expansion. In three dimensions, these results give $\sigma=\SIGMATHREE$ and $\nu_c=\NUCTHREE$. We locate the zero of the edge exponent at $d_\sigma=\DSIGMA$, also consistent with the six-loop estimate. Continuing the fixed point through $d_\sigma$, we obtain $\eta=-0.795(21)$ and $\omega=1.90(6)$ in $d=2$, compared with the exact values $-4/5$ and $2$.

  For the complex Yang--Lee fixed point, convergence of the field expansion depends crucially on its reference point. Expansions about $\phi=0$ fail to converge with the smooth regulator used here, whereas expansions about the running imaginary field value at which the mass vanishes converge down to $d=1.3$. This representation preserves the exact relation between the thermal eigenvalue and the field dimension, and hence the hyperscaling identity $\theta_h+\theta_t=d$, at every field-truncation order. We trace the large violations reported in earlier works to the omission of the expansion-point motion when linearizing the flow, rather than to the truncation itself.
\end{abstract}

\maketitle

\section{Introduction}
\label{sec:intro}

The Yang--Lee edge (YLE) arises from the zeros of a ferromagnetic partition function in a complex magnetic field. In the thermodynamic limit, these zeros accumulate along lines \cite{Yang:1952be,Lee:1952ig}, which, above the critical temperature, terminate at $h=\pm i h_c(T)$. The density of zeros becomes singular at these endpoints, and each endpoint is itself a critical point \cite{Kortman:1971zz,Fisher:1978pf}. Its long-distance behavior is described by a scalar field theory with an imaginary cubic interaction,
\begin{equation}
  \mathcal{L}=\tfrac12(\pd\phi)^2+i\,g\,\phi^3.
\end{equation}
The upper critical dimension of this theory is six. Below six dimensions, the field has a negative anomalous dimension $\eta$. The equation of motion relates $\phi^2$ to $\phi$ through a shift of the field, leaving $\eta$ as the only independent critical exponent \cite{Fisher:1978pf,Cardy:1985yy}. Interest in this universality class has recently been renewed by its role in studies of the QCD phase diagram: the location of the edge in the complex chemical-potential plane constrains the critical end point \cite{Mukherjee:2019eou,Bryant:2024twd,Clarke:2024ugt,Wan:2024xeu,Basar:2023nkp,Ding:2026fgp}. For an introduction to this connection, see Ref.~\cite{Skokov:2024fac}.

The edge exponents have been studied using the $\eps=6-d$ expansion, now available to six loops \cite{Macfarlane:1974vp,deAlcantaraBonfim:1980pe,deAlcantaraBonfim:1981sy,Gracey:2015tta,Kompaniets:2021hwg,Borinsky:2021jdb,Gracey:2025rnz}, the exact solution in $d=2$ \cite{Cardy:1985yy}, high-temperature series \cite{Kurtze:1979zz,Butera:2012tq}, the conformal bootstrap \cite{Gliozzi:2013ysa,Gliozzi:2014jsa}, lattice simulations \cite{LaiFisher1995,Hsu2005}, and fuzzy-sphere regularizations \cite{EliasMiro:2025msj,ArguelloCruz:2025zuq,Qian:2026fbx}. The functional renormalization group (FRG) \cite{Wetterich:1992yh,Morris:1993qb,Berges:2000ew,Pawlowski:2005xe,Delamotte:2007pf,Dupuis:2020fhh} offers a complementary non-perturbative approach that can be applied directly in both integer and non-integer dimensions~\cite{Ballhausen:2003gx,Codello:2012ec,Codello:2014yfa,Hellwig:2015woa,Defenu:2017dec,Yabunaka:2017uox,Chlebicki:2020pvo}. Examples include the Potts model, whose cubic interaction is closely related to the Yang--Lee case~\cite{BenAliZinati:2017vjy}, and long-range models, whose exponents follow from short-range ones at an effective fractional dimension~\cite{Defenu:2014bea}. At fourth and sixth order in the derivative expansion, it yields the exponents of the three-dimensional Ising and $O(N)$ universality classes in agreement with the conformal bootstrap, with error estimates that are under control \cite{Balog:2019rrg,DePolsi:2020pjk}. Recent developments include the use of conformal invariance to constrain truncations and the choice of regulator~\cite{Delamotte:2015aaa,Balog:2020fyt}; field reparametrisations that reduce the flow to its essential couplings or expand it about the physical ground state~\cite{Baldazzi:2021ydj,Ihssen:2023nqd,Ihssen:2024ihp}; and real-time and spectral formulations, which give access to dynamic critical exponents~\cite{Duclut:2016jct,Roth:2023wbp,Roth:2024rbi} and to the critical scaling of spectral functions~\cite{Braun:2022mgx,Horak:2023hkp,Kockler:2025kdt}.

Few previous FRG studies focused on the edge exponents.\footnote{The FRG has also been used to determine the universal \emph{location} of the YLE singularity \cite{Connelly:2020gwa,Rennecke:2022ohx,Johnson:2022cqv,Ihssen:2022xjv}.} An, Mesterh\'azy and Stephanov (AMS) \cite{An:2016lni} used a derivative expansion with the exponential regulator $R_k(q)=\alpha Z_kq^2/(e^{q^2/k^2}-1)$ at $\alpha=1$. They expanded the coefficient functions about the imaginary field value at which the mass vanishes. This expansion point runs with the FRG scale $k$. Their truncation included one of the three independent four-derivative operators. They reported a violation of an exact hyperscaling relation between two eigenvalues of the linearized flow: 15\% in $d=5$ and 60--70\% in $d=4$. Zambelli and Zanusso (ZZ) \cite{Zambelli:2016cbw} worked at second order in the derivative expansion with the Litim regulator $R_k(q)=aZ_k(k^2-q^2)\theta(k^2-q^2)$. They represented the field dependence by polynomials about $\phi=0$. These expansions converged slowly, and no fixed point was found below $d=4$.
Benedetti, Eustachon and Zanusso (BEZ) \cite{Benedetti:2026tpa} solved the LPA and LPA$'$ with the Litim regulator for a purely imaginary potential, by numerical shooting rather than polynomial expansion. Within the LPA$'$, they followed the fixed point through the dimension $d_0\simeq2.725$ at which $\Dphi$ vanishes, down to $d=2$. They obtained $\Dphi=0.167$, i.e.\ $\eta=-0.667$, in $d=3$.
Most recently, Beretta, S\'anchez-Villalobos and Codello \cite{Beretta:2026LY} applied functional dimensional regularization (FDR) \cite{Beretta:2026zcy}, a new functional approach, at second order in the derivative expansion; their values of $\eta$ are close to our $\ord{2}$ results (Sec.~\ref{sec:results}).

These studies leave two issues unresolved: the origin of the hyperscaling violation reported by AMS and the convergence of the polynomial expansion at lower dimensions. We revisit both issues here at $\ord{4}$. We also determine $\eta$ and $\sigma$ at fourth order in the derivative expansion for $1\le d\le6$, and locate the dimension $d_\sigma$ at which the edge exponent $\sigma$ changes sign,
\begin{align}
  \label{eq:dSigmaDef}
  \sigma(d_\sigma)=0\,.
\end{align}


For the YLE complex-valued fixed point, the convergence of the field expansion depends crucially on the choice of expansion point. With a suitable choice, we obtain convergent expansions down to $d \approx 1.3$, overcoming the convergence problem encountered by ZZ. This allowed us to find $d_\sigma=\DSIGMA$. This result agrees with the six-loop $\eps$ expansion \cite{Gracey:2025rnz,GraceyPrivate}.

The rest of the paper is organized as follows. Section~\ref{sec:constraints} summarizes the exact relations among the exponents. We introduce the FRG and the derivative expansion in Sec.~\ref{sec:setup}, derive the flow equations in Sec.~\ref{sec:flows}, and describe the representation of the field dependence in Sec.~\ref{sec:numerics}. Tests of the method are presented in Sec.~\ref{sec:validation}, followed by the results in Sec.~\ref{sec:results} and our conclusions in Sec.~\ref{sec:discussion}. Technical details are given in the appendices.

\section{Exponents and exact relations}
\label{sec:constraints}

We denote the scaling dimension of the field by $\Dphi=(d-2+\eta)/2$. The Yang--Lee fixed point has two relevant directions. The magnetic field $h$, which couples linearly to $\phi$, has eigenvalue/scaling dimension
\begin{equation}
  \theta_h=d-\Dphi=\frac{d+2-\eta}{2}.
\end{equation}
The second, ``thermal'' direction corresponds to the $\phi^2$ deformation of the cubic theory. The equation of motion, $\Box\phi\propto i\phi^2$, relates this deformation to a constant shift of the field, $\phi\to\phi+c$. Its eigenvalue is therefore the scaling dimension of $c$,
\begin{equation}
  \theta_t=\Dphi=\frac{d-2+\eta}{2},
  \label{eq:thetat}
\end{equation}
so that $\theta_h+\theta_t=d$ \cite{Fisher:1978pf,Cardy:1985yy}. The edge exponent and the correlation-length exponent follow as
\begin{equation}
  \sigma=\frac{\theta_t}{\theta_h}
  =\frac{d-2+\eta}{d+2-\eta},
  \qquad
  \nu_c=\frac{1}{\theta_h}
  =\frac{2}{d+2-\eta}.
  \label{eq:sigma}
\end{equation}

Near six dimensions, the anomalous dimension is $\eta=-\eps/9-43\eps^2/729+\dots$ \cite{Macfarlane:1974vp,deAlcantaraBonfim:1980pe,Borinsky:2021jdb}. In two dimensions, the exact solution gives $\eta=-4/5$, $\Dphi=-2/5$ and $\sigma=-1/6$ \cite{Cardy:1985yy}, and in one dimension $\sigma=-1/2$, i.e.\ $\eta=-1$ and $\Dphi=-1$ \cite{Fisher:1978pf}. Since $\sigma$ is positive in $d=3$ and negative in $d=2$, it must vanish at an intermediate dimension $d_\sigma\in(2,3)$. At this dimension, $\Dphi=0$ and $\eta=2-d_\sigma$.

In the FRG calculation, the two eigenvalues are extracted from the stability matrix of the flow linearized about the fixed point. The relations $\theta_h=d-\Dphi$ and $\theta_t=\Dphi$ are exact properties of the full flow, but they have different sensitivities to approximations of the field dependence.

The first relation follows from the fact that the flow depends on the potential $U(\phi)$ only through $U''$ and higher derivatives (see Sec.~\ref{sec:setup}). The coefficient of the linear term therefore decouples, giving $\theta_h=d-\Dphi$ exactly in \emph{any} truncation.

The second relation follows by considering an infinitesimal field shift in every fixed-point coefficient function, $F(\phi)\to F(\phi+c)$. The resulting perturbation is proportional to $F'$ and has eigenvalue $\theta_t=\Dphi$. Unlike the magnetic eigenvalue relation, this identity can be violated when the field dependence is approximated, for example by a truncated Taylor expansion. We use the relative discrepancy
\begin{equation}
  \eps_{\rm scale} =\frac{|\theta_t^{\rm stab}-\Dphi|}{\Dphi},
  \label{eq:eps_scale}
\end{equation}
where $\theta_t^{\rm stab}$ is the stability-matrix eigenvalue, to quantify such violations. Comparing this eigenvalue with $\Dphi$ obtained from $\eta$ provides a direct check on the treatment of the field dependence.

\section{FRG and the derivative expansion}
\label{sec:setup}

The FRG describes the evolution of the scale-dependent effective action $\Gamma_k[\phi]$. Fluctuations with momenta $q\gtrsim k$ are integrated out, while those with $q\lesssim k$ are suppressed by a regulator term $\tfrac12\int\phi\,R_k\,\phi$, which acts as a mass of order $k^2$. The flow interpolates between the microscopic action at $k=\Lambda$ and the full effective action as $k\to0$. It obeys the exact Wetterich equation \cite{Wetterich:1992yh},
\begin{equation}
  \pd_t\Gamma_k =\frac12\,\mathrm{Tr}\Big[ \pd_tR_k\,\big(\Gamma_k^{(2)}+R_k\big)^{-1} \Big],
  \qquad t=\ln(k/\Lambda),
  \label{eq:wetterich}
\end{equation}
where $\Gamma_k^{(2)}$ denotes the second functional derivative. Although this equation has a one-loop form, it contains the full, field-dependent propagator and vertices. In dimensionless variables, a critical point corresponds to a fixed point of the flow.

We solve Eq.~\eqref{eq:wetterich} by performing a derivative expansion of $\Gamma_k$. At leading order, the local potential approximation (LPA) retains a running potential $U(\phi)$ and a fixed kinetic term. At second order, $\ord{2}$, the kinetic coefficient is promoted to a field-dependent wave-function renormalization $Z(\phi)$. At fourth order, the ansatz is
\begin{multline}
  \Gamma_k=\int\!\dd^dx\Big\{ U(\phi)+\tfrac12Z(\phi)(\pd\phi)^2 +\tfrac12W_1(\phi)(\Box\phi)^2\\ +\tfrac12W_2(\phi)(\pd\phi)^2\Box\phi +\tfrac12W_3(\phi)\big[(\pd\phi)^2\big]^2 \Big\}.
  \label{eq:ansatz}
\end{multline}
Up to total derivatives, this ansatz contains all independent operators with at most four derivatives. In the Ising model, $\mathbb{Z}_2$ symmetry requires $W_2$ to be odd in $\phi$. At the Yang--Lee edge, this symmetry is explicitly broken, and we retain all three functions $W_{1,2,3}$. To distinguish the complete fourth-order expansion from the truncation used by AMS, which retains only $W_1$, we denote them by $\ord{4}$ and $\ord{4}_{W_1}$, respectively.

For the regulator, we use $R_k(q)=Z_0k^2\,r(q^2/k^2)$ with the smooth exponential shape
\begin{equation}
  r(y)=\alpha\,\frac{y}{e^y-1},
  \label{eq:regulator}
\end{equation}
as in Refs.~\cite{An:2016lni,Canet:2003qd,Balog:2019rrg}. We write $Z(\phi)=Z_0z(\phi)$ and impose $z(\phi_0)=1$ at a reference field $\phi_0$, whose choice is specified below. Varying the amplitude $\alpha$ allows us to assess the residual regulator dependence. This dependence would be absent in an exact solution of the Wetterich equation, but remains at finite truncation order and provides a useful diagnostic of the truncation.

The Litim regulator \cite{Litim:2000ci,Litim:2001up} is not sufficiently smooth for the derivative expansion beyond $\ord{2}$, since the $\ord{4}$ flows require higher momentum derivatives of $R_k$. One may instead introduce terms that are non-analytic in momentum, but the resulting proliferation of coefficient functions makes this regulator impractical for the higher-order calculations considered here.

We introduce dimensionless fields and coefficient functions by rescaling with appropriate powers of $k$ and $Z_0$, and define the anomalous dimension as $\eta=-\pd_t\ln Z_0$. For a coefficient function $f$ multiplying an operator with $2s$ derivatives and $n$ explicit fields, the flow takes the form
\begin{multline}
  \pd_tf= \Big(-d+2s+\tfrac{n}{2}(d-2+\eta)\Big)f +\Dphi\,\phi f'\\ +\int_0^\infty\!\dd y\,y^{d/2-1}\,I_f(y),
  \label{eq:flow}
\end{multline}
where $y=q^2/k^2$. The first two terms describe the scaling of the operator and the field argument. The integral contains the loop contribution, whose integrand depends on the coefficient functions, their field derivatives, the regulator, $d$ and $\eta$. A fixed point is a set of functions for which all flows $\pd_tf$ vanish.

Although the Yang--Lee fixed point is complex, it is invariant under $\phi\to-\phi$ combined with complex conjugation. The coefficient functions therefore satisfy $F(-\phi)^*=\pm F(\phi)$, with the minus sign applying to $W_2$ because its operator is odd under field reflection. For an expansion about a point on the imaginary axis, this symmetry constrains successive Taylor coefficients to alternate between real and purely imaginary values. Factoring out the corresponding powers of $i$ turns the fixed-point equations into a real system.

\section{Deriving the flow equations}
\label{sec:flows}

We derive the flows of $U$, $Z$ and $W_{1,2,3}$ by matching the momentum dependence of the vertex functions at a constant background field $\phi$. Here $\Gamma^{(n)}$ denotes the $n$-th functional derivative of $\Gamma_k$ in momentum space. The ansatz~\eqref{eq:ansatz} expresses each vertex as a polynomial in the external momenta $p_1,\dots,p_n$, with coefficients determined by the coefficient functions and their field derivatives. For the two-point function, for example,
\begin{equation}
  \Gamma^{(2)}(p)=U''+Zp^2+W_1p^4.
\end{equation}
Taking $n$ functional derivatives of the Wetterich equation~\eqref{eq:wetterich} gives a second expression for $\pd_t\Gamma^{(n)}$: a sum of one-loop diagrams containing the propagator $(\Gamma^{(2)}+R_k)^{-1}$ and the vertices $\Gamma^{(3)},\dots,\Gamma^{(n+2)}$ obtained from the ansatz.

We expand these loop expressions to fourth order in the external momenta and match the coefficients of the resulting momentum structures. The flows are extracted as follows:
\begin{itemize}
  \item $\pd_tU$ from the Wetterich equation evaluated at a constant field;
  \item $\pd_tZ$ and $\pd_tW_1$ from the $p^2$ and $p^4$ coefficients of the two-point flow;
  \item $\pd_tW_2$ and $\pd_tW_3$ from the three- and four-point flows, respectively.
\end{itemize}
We perform the angular averages over the loop momentum analytically for arbitrary $d$. Each flow then reduces to a one-dimensional integral over $y=q^2/k^2$, of the form given in Eq.~\eqref{eq:flow}. These steps are implemented in Mathematica for a general operator basis, as described in Appendix~\ref{app:derivation}.

\subsection{Overdetermined projections}

Matching the momentum structures yields more equations than unknown flows. For the three-point function $\Gamma^{(3)}(p_1,p_2,p_3)$, with $p_1+p_2+p_3=0$, the expansion through fourth order contains 10 momentum structures. Their coefficients involve only four unknown flows: $\pd_tW_2$, $\pd_tW_1'$, $\pd_tZ'$ and $\pd_tU'''$. The four-point function similarly yields 28 equations for five unknowns.

These additional equations provide consistency checks. Through fourth order in the external momenta, the loop expansion has the momentum dependence of a local functional that can be expressed in the basis~\eqref{eq:ansatz}. The coefficient equations must therefore be mutually consistent. In addition, the flows of $Z'$, $W_1'$ and the other field derivatives extracted from higher-point vertices must agree with the field derivatives of the flows obtained from lower-point vertices.

We verify that the residuals of all redundant equations vanish identically and that the field-derivative relations hold for all $y$. These checks test the vertex algebra, the diagrammatic combinatorics and the angular averages before numerical integration. The resulting integrands are exported as Julia code, which is used throughout the numerical calculations.

\section{Representing the field dependence}
\label{sec:numerics}

The fixed-point equations form a set of coupled ordinary differential equations in the field for $U$, $Z$ and $W_i$. We solve them by expanding each function about a reference point $\phi=i\chi$ on the imaginary axis,
\begin{equation}
  F(\phi)=\sum_{n=0}^{N_F}\frac{f_n}{n!}(\phi-i\chi)^n,
  \label{eq:taylor}
\end{equation}
with $N_U=N_Z+3$, $N_{W_1}=N_{W_3}=N_Z-2$ and $N_{W_2}=N_Z-1$. These relative orders follow from the structure of the flow equations (Appendix~\ref{app:fieldconv}). We solve the resulting equations for the Taylor coefficients by Newton's method, starting from the perturbative fixed point near $d=6$ and continuing in small steps in $d$ towards $d=1$. The stability matrix is obtained by automatic differentiation of the same code, and its eigenvalues are then computed. We check convergence in the field expansion by increasing $N_Z$.

\begin{figure}[t]
  \centering
  \includegraphics[width=\columnwidth]{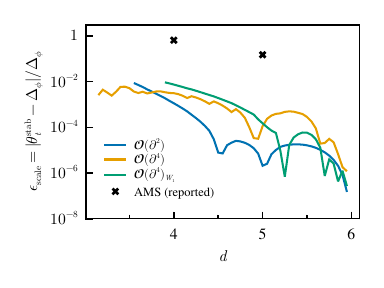}
  \caption{Violation $\eps_{\rm scale}$ of the exact relation $\theta_t=\Dphi$ for expansions about fixed imaginary points. Each curve ends where the field expansion fails. For the running massless point, the violation vanishes identically, $\eps_{\rm scale} \equiv 0$. Crosses show the violations reported in Ref.~\cite{An:2016lni}.}
  \label{fig:diag}
\end{figure}

Convergence depends strongly on the choice of $\chi$. The fixed-point functions have singularities in the complex field plane associated with zeros of the regulated inverse propagator, $U''+Zq^2+\dots+R_k$. Their distance from the expansion point limits the radius of convergence of Eq.~\eqref{eq:taylor}. We compare three choices of reference point.

\begin{enumerate}
  \item \emph{Expansion about $\phi=0$.}
    The origin is a natural choice for a $\mathcal{PT}$-symmetric problem and was used in Ref.~\cite{Zambelli:2016cbw}. With the exponential regulator, however, this expansion does not converge. At $d=5.5$, the anomalous dimension oscillates with period three as $N_Z$ increases (Fig.~\ref{fig:fieldconv}), indicating singularities at angles of approximately $120^\circ$ around the origin. We therefore do not pursue this choice further.

  \item \emph{Fixed imaginary point.}
    Choosing $\chi=\chi_0<0$ moves the expansion point away from the nearest singularity on the positive imaginary axis and substantially improves convergence. At $d=5.5$, $\eta$ is stable to seven digits for $N_Z\ge10$ (Fig.~\ref{fig:fieldconv}). As we lower $d$, we choose $\chi_0(d)=-2-\max(5.5-d,0)$ and impose $z(i\chi_0)=1$. The expansion point is thus fixed along each RG flow, but varies between calculations at different dimensions.

    Where it converges, this representation gives the fastest convergence, with $\eps_{\rm scale}<10^{-2}$ (Fig.~\ref{fig:diag}). Below $d\simeq3.2$, a pair of off-axis singularities approaches the imaginary axis and this choice of expansion point ceases to work. A different choice of $\chi_0$ might extend its range, but the running-point representation described next avoids the need for such an adjustment.

  \item \emph{Running massless point.}
    Following Refs.~\cite{An:2016lni,Rennecke:2022ohx,Johnson:2022cqv}, we expand about the point on the imaginary axis where the mass vanishes and normalize $z$ at the same point:
    \begin{equation}
      U''(i\chi)=0,\qquad z(i\chi)=1.
      \label{eq:massless}
    \end{equation}
    The position $\chi$ now runs with $k$. The flows of the Taylor coefficients at the massless point depend on it only through the combination $\psi=i(\pd_t\chi+\Dphi\chi)$, see Appendix~\ref{app:translation}. The two conditions~\eqref{eq:massless} determine $\psi$ and $\eta$. We therefore solve for the Taylor coefficients, $\eta$ and $\psi$; the position itself can be found through  $\pd_t\chi=-\Dphi\chi-i\psi$, however, it does not feed back into the other flows. This choice of the expansion point has two useful properties.

    \begin{itemize}
      \item \emph{The translation relation is preserved exactly.}
        A translation of the fixed-point solution shifts the massless point but leaves the Taylor coefficients evaluated there \emph{unchanged}. The corresponding perturbation is therefore the unit vector in the $\chi$ direction. It is an eigenvector of the stability matrix with eigenvalue $\Dphi$ at every truncation order. Because $\chi$ does not feed back into the other flows, this eigenvalue is exact, and the remaining eigenvalues are those of the stability matrix of the Taylor coefficients alone. The explicit $\chi$ dependence of the dilatation term $\Dphi\phi F'$ is cancelled by the motion of the expansion point; Appendix~\ref{app:translation} gives the explicit derivation. Consequently, $\eps_{\rm scale}=0$ \emph{identically} in this representation.

      \item \emph{The massless point recedes as $\Dphi\to0$.}
        At $U''(i\chi)=0$, only the dilatation term $i\Dphi\chi\,U'''(i\chi)$ and the loop contribution remain in the fixed-point equation for $U''$ (see Eq.~\eqref{eq:flow}). Denoting the loop contribution by $L_2$, we obtain
        \[
          i\Dphi\,\chi\,U'''(i\chi)=-L_2.
        \]
        Both $U'''$ and $L_2$ remain finite as $d$ varies, so the product $\chi\Dphi$ is finite and $\chi\propto1/\Dphi$. As $d$ approaches $d_\sigma$ and $\Dphi$ vanishes, the massless point moves toward $-i\infty$, away from the singularities.
    \end{itemize}

    The drawback is slower convergence in $N_Z$ at $\ord{4}$ near six dimensions.
\end{enumerate}

We choose the expansion point according to its convergence. At $\ord{2}$, we use the running massless point throughout, with convergence in $\eta$ better than $5\times10^{-4}$ as $N_Z$ is increased for $d\ge d_\sigma$. At $\ord{4}$, we use fixed imaginary points for $d\ge3.2$, where they give more accurate results, and the running massless point below this dimension, down to $d=1$.

The two choices of the expansion point agree within $2\times10^{-4}$ for $d\ge4$ and within $3\times10^{-3}$ at $d\simeq3.4$. Because they impose the normalization of $z$ at different field values, we include their difference in the uncertainty estimate (Appendix~\ref{sec:errors}).

\section{Tests: the three-dimensional Ising model,
the six-dimensional limit and the AMS calculation}
\label{sec:validation}

We test the calculation against three benchmarks: the three-dimensional Ising fixed point, the perturbative Yang--Lee result near six dimensions, and the results obtained with the AMS truncation.

\paragraph*{Three-dimensional Ising universality.}
Our flow equations were derived without imposing a field symmetry. The Ising fixed point, for which $\mathbb{Z}_2$ symmetry requires $U$, $Z$, $W_1$ and $W_3$ to be even and $W_2$ to be odd, therefore tests the derivation in a different symmetry setting. For this benchmark, we resolve the field dependence on a grid (Appendix~\ref{app:collocation}) and determine the stationary points of the exponents as functions of $\alpha$, following the principle of minimal sensitivity (PMS) \cite{Canet:2002gs}.

Table~\ref{tab:ising} compares our results with those of Ref.~\cite{Balog:2019rrg}, which used the same regulator family. The LPA values of $\nu$ differ by $10^{-5}$, and the $\ord{4}$ values of $\eta$ by $3\times10^{-5}$. The largest difference in the table is $7.4\times10^{-4}$. These small differences are consistent with the use of different normalization points and PMS prescriptions.

\begin{table}[t]
  \caption{Three-dimensional Ising exponents obtained with $r(y)=\alpha y/(e^y-1)$ at their stationary points in $\alpha$, compared with Ref.~\cite{Balog:2019rrg} for the same regulator family. For comparison, Ref.~\cite{Canet:2003qd} gives $\nu=0.6281$ and $\eta=0.0443$ at $\ord{2}$, and $\nu=0.632$ and $\eta=0.033$ at $\ord{4}$.}
  \label{tab:ising}
  \begin{ruledtabular}
    \begin{tabular}{lllll}
      & \multicolumn{2}{c}{This work}
      & \multicolumn{2}{c}{Ref.~\cite{Balog:2019rrg}}\\
      & $\nu$ & $\eta$ & $\nu$ & $\eta$\\\hline
      LPA & 0.65060 & 0 & 0.65059 & 0\\
      $\ord{2}$ & 0.62787 & 0.04426 & 0.62779 & 0.04500\\
      $\ord{4}$ & 0.62997 & 0.03457 & 0.63027 & 0.03454
    \end{tabular}
  \end{ruledtabular}
\end{table}

\paragraph*{Approach to six dimensions.}
As $\eps=6-d\to0$, the Yang--Lee fixed point approaches the perturbative fixed point of the cubic theory. A truncation that includes $Z(\phi)$ must reproduce the leading result, $\eta=-\eps/9+O(\eps^2)$. Fitting $\eta(\eps)$ over $\eps\in[0.005,0.04]$, we obtain the limiting ratios $\eta/\eps\to-0.1110$ at $\ord{2}$ and $-0.1112$ at $\ord{4}$, both in agreement with $-1/9$.

\paragraph*{Reproduction of the AMS calculation.}
AMS used the running massless point~\eqref{eq:massless}, the regulator amplitude $\alpha=1$, and the polynomial truncation $(n_U,n_Z,n_W)=(7,5,0)$. Their notation specifies the highest power in $U$ by $n_U$, while $n_Z$ and $n_W$ count the terms retained in $Z$ and $W_1$ [Eqs.~(14)--(15) of Ref.~\cite{An:2016lni}]. This corresponds to our $\ord{2}$ truncation with $(N_U,N_Z)=(7,4)$. With these settings, we obtain $\eta=-0.1260$ in $d=5$ and $-0.3156$ in $d=4$, reproducing the AMS values $-0.126(6)$ and $-0.316(16)$ \cite{An:2016lni}.

For the same truncation, we find $\theta_t^{\rm stab}=\Dphi$ to machine precision, as required by the argument in Sec.~\ref{sec:numerics}. The violations of 15\% and 60--70\% reported by AMS therefore do not arise from the truncation. They instead result from the treatment of the running expansion point when linearizing the flow, as discussed in Sec.~\ref{sec:addendum}.

This issue does not affect the exponents that AMS obtained from $\eta$ using Eq.~\eqref{eq:sigma}. Their $\ord{4}$ results are, however, affected by the omission of operators from the fourth-order truncation, which we examine in Sec.~\ref{sec:omitted}.

\section{Results}
\label{sec:results}

\subsection{Anomalous dimension and exponents}
\label{sec:resO2}

\begin{figure*}[t]
  \centering
  \includegraphics[width=\textwidth]{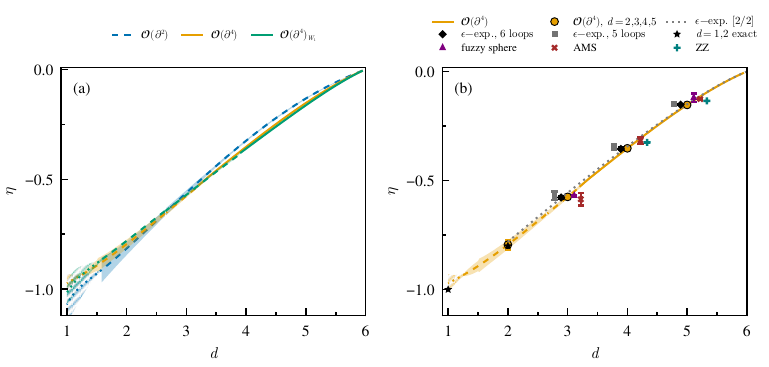}
  \caption{Anomalous dimension of the Yang--Lee edge as a function of $d$ at $\alpha=2$.
    (a) Comparison of the $\ord{2}$, complete $\ord{4}$ and $\ord{4}_{W_1}$ truncations.
  (b) Complete $\ord{4}$ result compared with the literature. Filled circles mark our values at $d=2,\ 3$, $4$ and $5$. Above $d_\sigma$, bands combine the field-truncation, regulator and normalization-point uncertainties, while below $d_\sigma$, only the first two. Solid lines use fixed imaginary expansion points; dashed lines use the running massless point (Sec.~\ref{sec:numerics}). The $\ord{2}$ curve uses the running massless point throughout and is dashed over the whole range. Where the field expansion is not converged (at $\ord{4}$ below $d\simeq1.3$, at $\ord{2}$ and $\ord{4}_{W_1}$ below $d\simeq1.6$), the lines are dotted and the hatched bands show the change from $N_Z=14$ to $16$ ($12$ to $14$ at $\ord{2}$ and $\ord{4}_{W_1}$). The dotted gray line in (b) is the $[2/2]$ Pad\'e approximant of the five-loop $\eps$ expansion \cite{Borinsky:2021jdb}. Other symbols show six-loop \cite{Gracey:2025rnz} and five-loop \cite{Borinsky:2021jdb} $\eps$-expansion estimates, the exact results in $d=2$ and $d=1$ \cite{Cardy:1985yy,Fisher:1978pf}, fuzzy-sphere results \cite{ArguelloCruz:2025zuq,Qian:2026fbx}, AMS \cite{An:2016lni}, and ZZ \cite{Zambelli:2016cbw}. All literature values are at integer $d$; symbols are displaced horizontally for visibility.}
  \label{fig:eta}
\end{figure*}

\begin{figure}[t]
  \centering
  \includegraphics[width=\columnwidth]{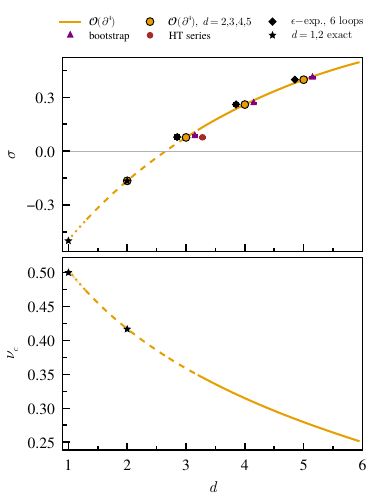}
  \caption{Edge exponent $\sigma$ and correlation-length exponent $\nu_c$, obtained from Eq.~\eqref{eq:sigma} using the complete $\ord{4}$ result in Fig.~\ref{fig:eta}(b). Filled circles mark our values of $\sigma$ at $d=2,\ 3$, $4$ and $5$; the error bar at $d=2$ is the total uncertainty. As in Fig.~\ref{fig:eta}, the lines are solid where fixed imaginary expansion points are used ($d\ge3.2$) and dashed where the running massless point is used; they are dotted below $d\simeq1.3$, where the field expansion is not converged. Other symbols show results from the six-loop $\eps$ expansion \cite{Gracey:2025rnz}, conformal bootstrap \cite{Gliozzi:2014jsa}, high-temperature series \cite{Butera:2012tq}, and the exact results in $d=2$ and $d=1$ \cite{Cardy:1985yy,Fisher:1978pf}, shown as stars for both $\sigma$ and $\nu_c$. Literature values are at integer $d$; symbols are displaced horizontally for visibility.}
  \label{fig:sigma}
\end{figure}

Figure~\ref{fig:eta}(a) compares $\eta(d)$ in the three truncations, while Fig.~\ref{fig:eta}(b) compares the complete $\ord{4}$ result with earlier determinations. The corresponding edge and correlation-length exponents, $\sigma$ and $\nu_c$, are shown in Fig.~\ref{fig:sigma}. Table~\ref{tab:exponents} lists the results at selected dimensions. Throughout this section, we use the regulator amplitude $\alpha=2$; the reasons for this choice are discussed in Sec.~\ref{sec:regulator}.

Going from $\ord{2}$ to the complete $\ord{4}$ truncation changes $\eta$ by about 16\% in $d=5$, 9\% in $d=4$ and 3\% in $d=3$. Our fourth-order results are
\begin{equation}
  \eta=
  \begin{cases}
    \ETAOFOURFIVETOT, & d=5,\\
    \ETAOFOURFOURTOT, & d=4,\\
    \ETAOFOURTHREETOT, & d=3.
  \end{cases}
\end{equation}
The quoted uncertainties include the derivative-expansion estimate described in Appendix~\ref{sec:errors}. In $d=3$, these values give $\sigma=\SIGMATHREE$ and $\nu_c=\NUCTHREE$.

The results agree with the six-loop $\eps$-expansion analysis of Ref.~\cite{Gracey:2025rnz}, which gives $\eta=-0.152(1)$, $-0.355(6)$ and $-0.578(6)$ in $d=5$, $4$ and $3$, respectively, and $\sigma=0.078(9)$ in $d=3$. Our three-dimensional result also agrees with the fuzzy-sphere determination $\Dphi=0.2151(8)$, corresponding to $\eta=-0.570(2)$ \cite{ArguelloCruz:2025zuq}, and with the high-temperature series estimate $\sigma=0.077(2)$ \cite{Butera:2012tq}. The conformal-bootstrap estimate, $\sigma=0.085(1)$ \cite{Gliozzi:2014jsa}, has a larger central value.

At $\ord{2}$, we find close agreement with ZZ \cite{Zambelli:2016cbw}, who used the Litim regulator. Our values, $\eta=\ETAOTWOFIVE$ in $d=5$ and $\ETAOTWOFOUR$ in $d=4$, compare with their $-0.1344(1)$ and $-0.325(3)$, respectively. The FDR values at the same order, $\eta=-0.1414$, $-0.3276$ in $d=5$, $4$~\cite{Beretta:2026LY} agree with ours at $\ord{2}$; their values at $d=3$ is $-0.5451$ vs our  $\ETAOTWOTHREE$.

\begin{table*}[t]
  \caption{Anomalous dimension $\eta$ at $\alpha=2$ at second and fourth order of the derivative expansion. The $\ord{2}$ results use the running massless point. The $\ord{4}$ results use fixed imaginary points for $d\ge3.2$ and the running massless point below this dimension (Sec.~\ref{sec:numerics}). Parenthetical errors combine the regulator uncertainty (half the spread over $\alpha\in[1,3]$, or the shift to the PMS point if larger), the field-order difference and the normalization-point dependence. For $\ord{4}$, the second error, given after $\pm$, is the derivative-expansion estimate $|\eta_{\ord{4}}-\eta_{\ord{2}}|/3$ (Appendix~\ref{sec:errors}). The exponents $\sigma$ and $\nu_c$ follow from Eq.~\eqref{eq:sigma} using the $\ord{4}$ value of $\eta$. The final column gives the six-loop $\eps$-expansion estimate of Ref.~\cite{Gracey:2025rnz}. Below $d_\sigma$ (lower block), both truncations use the running massless point; the errors combine the regulator and field-order uncertainties, and the final column gives the exact values \cite{Cardy:1985yy,Fisher:1978pf}. Entries marked $\dagger$ are not converged in the field expansion (Sec.~\ref{sec:lowd}); for these values of $\eta$, the parenthetical error is the change from $N_Z=14$ to $16$ ($12$ to $14$ at $\ord{2}$).}
  \label{tab:exponents}
  \begin{ruledtabular}
    \begin{tabular}{rlllll}
      $d$ & $\eta\ [\ord{2}]$ & $\eta\ [\ord{4}]$
      & $\sigma$ & $\nu_c$
      & $\eta$ [$\eps$-exp./exact]\\\hline
      5.5 & $-0.0588(16)$ & $-0.06855(29)\pm0.003$
      & $0.4534$ & $0.2643$ & --\\
      5 & $-0.1324(27)$ & $-0.15307(52)\pm0.007$
      & $0.3980$ & $0.2796$ & $-0.152(1)$\\
      4.5 & $-0.2214(35)$ & $-0.24868(58)\pm0.009$
      & $0.3336$ & $0.2964$ & --\\
      4 & $-0.3244(38)$ & $-0.3523(19)\pm0.009$
      & $0.2594$ & $0.3148$ & $-0.355(6)$\\
      3.5 & $-0.4388(52)$ & $-0.4603(34)\pm0.007$
      & $0.1744$ & $0.3356$ & --\\
      3 & $-0.562(12)$ & $-0.5751(76)\pm0.005$
      & $0.0762$ & $0.3587$ & $-0.578(6)$\\\hline
      2.5 & $-0.688(17)$ & $-0.687(13)\pm0.001$
      & $-0.0360$ & $0.3856$ & --\\
      2 & $-0.814(30)$ & $-0.795(20)\pm0.007$
      & $-0.1657$ & $0.4171$ & $-0.8$ (exact)\\
      1.5 & $-0.93(6)^\dagger$ & $-0.893(23)\pm0.012$
      & $-0.3170$ & $0.4553$ & --\\
      1 & $-1.07(8)^\dagger$ & $-0.97(4)^\dagger\pm0.03$
      & $-0.4966^\dagger$ & $0.5034^\dagger$ & $-1$ (exact)\\
    \end{tabular}
  \end{ruledtabular}
\end{table*}

\subsection{Effect of the omitted operators}
\label{sec:omitted}

Retaining only $W_1$, as in Ref.~\cite{An:2016lni}, gives $\eta(5)=-0.163$ and $\eta(4)=-0.361$. Measured relative to $\ord{2}$, these shifts are 1.5 and 1.3 times those of the complete $\ord{4}$ truncation in $d=5$ and $4$, respectively. The combined contribution of $W_2(\pd\phi)^2\Box\phi$ and $W_3[(\pd\phi)^2]^2$ therefore opposes the shift produced by $W_1$ alone and is not negligible.

The pattern changes near $d=3$. There, $\ord{4}_{W_1}$ gives $\eta(3)=-0.571$, between the $\ord{2}$ and complete $\ord{4}$ values. The discrepancy caused by omitting $W_2$ and $W_3$ thus varies in both magnitude and sign with dimension.

\subsection{Regulator dependence}
\label{sec:regulator}

\begin{figure*}[t]
  \centering
  \includegraphics[width=0.85\textwidth]{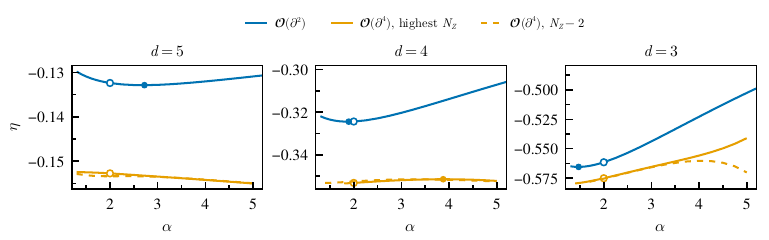}
  \caption{Regulator dependence of $\eta$ in $d=5$, $4$ and $3$, calculated with the running massless point. At $\ord{2}$, $N_Z=10$. At $\ord{4}$, solid lines show the highest field order computed: $N_Z=14$ in $d=5$ and $4$, and $N_Z=12$ in $d=3$. Dashed lines show the preceding field order. Filled circles mark stationary points $\alpha_{\rm PMS}$; open circles mark $\alpha=2$, used for the reported results. In $d=3$, the two $\ord{4}$ curves separate for $\alpha\gtrsim3$, where the field expansion is not converged.}
  \label{fig:pms}
\end{figure*}

At finite truncation order, the results depend on the regulator amplitude $\alpha$ in Eq.~\eqref{eq:regulator}. The principle of minimal sensitivity (PMS) selects an amplitude $\alpha_{\rm PMS}$ at which $\eta$ is stationary \cite{Canet:2002gs}. Figure~\ref{fig:pms} shows $\eta(\alpha)$ in $d=5$, $4$ and $3$, calculated with the running massless point.

At $\ord{2}$, each curve has a single extremum, at $\alpha_{\rm PMS}=\APMSOTWOFIVE$, $\APMSOTWOFOUR$ and $\APMSOTWOTHREE$ in $d=5$, $4$ and $3$, respectively. The corresponding values of $\eta$, $\EPMSOTWOFIVE$, $\EPMSOTWOFOUR$ and $\EPMSOTWOTHREE$, differ from the $\alpha=2$ results by at most $\DPMSMAXOTWO$.

At $\ord{4}$, a reliable search for a stationary point requires convergence of the field expansion over a sufficiently wide range of $\alpha$. With fixed imaginary expansion points, the variation of $\eta$ over $\alpha\in[1,3]$ is only $1\times10^{-3}$ in $d=5$ and $2.5\times10^{-3}$ in $d=4$. Larger amplitudes are accessible only with the running point, for which convergence is slower. We therefore extended those calculations to $N_Z=14$ (Fig.~\ref{fig:pms}).

In $d=5$, the $N_Z=12$ and $14$ results agree within $10^{-4}$ for $\alpha\ge3$, and no stationary point is found over $\alpha\in[1.3,5]$. In $d=4$, they agree within $6\times10^{-4}$ for $\alpha\ge2.5$ and exhibit a shallow maximum at $\alpha_{\rm PMS}\simeq3.3$--$3.9$. Its value, $\eta=\EPMSOFOURFOUR$, lies $1.7\times10^{-3}$ above the $\alpha=2$ result.

In $d=3$, the expansion is converged for $\alpha\lesssim3$, where the $N_Z=10$ and $12$ results agree within $3\times10^{-4}$. Maxima appear at $\alpha\simeq3.4$ for $N_Z=8$ and at $\alpha\simeq4.0$ for $N_Z=10$, but move to larger amplitudes as the field order increases and are absent at $N_Z=12$. They are therefore artifacts of the field truncation. Within the converged range, $\eta$ increases monotonically with $\alpha$ and has no stationary point.

We consequently use $\alpha=2$ for all dimensions and truncations. The regulator uncertainty is estimated as half the spread of $\eta$ over $\alpha\in[1,3]$, or as the shift to the PMS point when that is larger. The latter applies only at $\ord{4}$ in $d=4$ (Appendix~\ref{sec:errors}).

\subsection{Dimension at which
\texorpdfstring{$\sigma$}{sigma} vanishes}
\label{sec:dsigma}

The change in sign of $\sigma$ between $d=3$ and $d=2$ implies a zero at an intermediate dimension $d_\sigma$. Using the running massless point, we can follow the fixed point through this dimension: as $\Dphi\to0$, the expansion point recedes to infinity, but the flows of the Taylor coefficients do not depend on its position (Sec.~\ref{sec:numerics}).

\begin{figure}[t]
  \centering
  \includegraphics[width=\columnwidth]{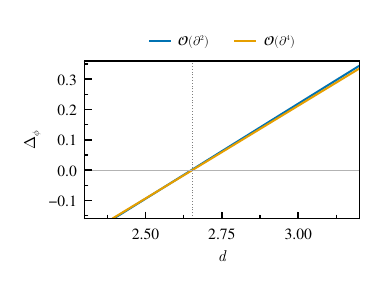}
  \caption{Field dimension $\Dphi$ through $d_\sigma$ at $\alpha=2$, at $\ord{2}$ and in the complete $\ord{4}$ truncation ($N_Z=16$). The dotted vertical line marks the zero of $\Dphi$ at $\ord{4}$, where $\sigma=0$.}
  \label{fig:dsigma}
\end{figure}

Figure~\ref{fig:dsigma} shows $\Dphi$ near $d_\sigma$. It passes through zero linearly. Increasing the field order to $N_Z=14$ and $16$ at $\ord{4}$ changes $d_\sigma$ by less than $5\times10^{-4}$.

We obtain $d_\sigma=\DSIGMAOTWO$ at $\ord{2}$ and $\DSIGMAOFOUR$ at $\ord{4}$. The uncertainties combine the field-order dependence and the regulator variation over $\alpha\in[1,3]$. Taking half the difference between the two orders as the derivative-expansion uncertainty gives
\begin{equation}
  d_\sigma=\DSIGMA.
\end{equation}
At this dimension,
$\eta=2-d_\sigma\simeq\ETADSIGMA$.

An independent estimate follows from the $\eps$ expansion. Applying the six-loop Pad\'e analysis of Ref.~\cite{Gracey:2025rnz} to $\sigma$ gives $d_\sigma=2.649(30)$ and $2.649(25)$ \cite{GraceyPrivate}. These estimates use, respectively, the mean and the loop-order-weighted mean of the valid Pad\'e approximants, and both agree with our result. A conformal-bootstrap calculation using the determinant method found $d_\sigma=2.6199$, without an uncertainty estimate \cite{Hikami:2017hwv}.

\subsection{Correction to scaling and $\omega$ exponent}
\label{sec:omega}

The leading correction to scaling\footnote{We thank C.~A.~S\'anchez-Villalobos for prompting us to look at $\omega$.} is governed by the smallest irrelevant eigenvalue of the stability matrix, $\omega=-\theta_3>0$~\footnote{$\theta_1=\theta_h$ and $\theta_2=\theta_t$ are the two relevant eigenvalues.}. Unlike $\theta_h$ and $\theta_t$, $\omega$ is independent of $\eta$.

\begin{figure}[t]
  \centering
  \includegraphics[width=\columnwidth]{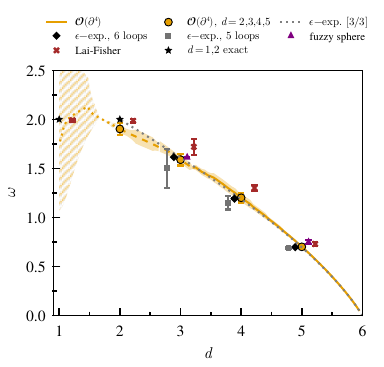}
  \caption{Correction-to-scaling exponent $\omega$ in the complete $\ord{4}$ truncation at $\alpha=2$. The band combines the field-truncation, regulator and normalization-point uncertainties; the solid line uses fixed imaginary expansion points and the dashed line the running massless point. Filled circles mark our values at $d=2,\ 3$, $4$ and $5$, with error bars that include the derivative-expansion estimate. Below $d=2$, where the field expansion of $\omega$ is not converged, the line is dotted and the hatched band shows the change from $N_Z=14$ to $16$. The dotted gray line is the $[3/3]$ Pad\'e approximant of the six-loop $\eps$ expansion of $\omega/\eps$ \cite{Gracey:2025rnz}, constrained by the exact value $\omega=2$ in $d=2$. Other symbols show the six-loop \cite{Gracey:2025rnz} and five-loop \cite{Borinsky:2021jdb} $\eps$-expansion estimates, the fuzzy-sphere results in $d=3$ \cite{ArguelloCruz:2025zuq} and $d=5$ \cite{Qian:2026fbx}, the estimates of Ref.~\cite{LaiFisher1995} (including $d=1$ and $2$), and the exact $d=1$ and $d=2$ values. Literature symbols are displaced horizontally for visibility.}
  \label{fig:omega}
\end{figure}

We compute $\omega$ with the same representations of the field dependence and the same choice $\alpha=2$ as for $\eta$.
The result is displayed in Figure~\ref{fig:omega}. Numerically, at $\ord{4}$ we find
\begin{equation}
  \omega=
  \begin{cases}
    0.701(8)\pm0.004, & d=5,\\
    1.20(4)\pm0.02, & d=4,\\
    1.59(3)\pm0.05, & d=3,
  \end{cases}
\end{equation}
where the second error is the derivative-expansion estimate $|\omega_{\ord{4}}-\omega_{\ord{2}}|/3$.
These values agree with the six-loop constrained-Pad\'e estimates of Ref.~\cite{Gracey:2025rnz}, $\omega=0.6970(2)$, $1.1924(10)$ and $1.615(1)$ in $d=5$, $4$ and $3$, respectively, and with the five-loop estimates $0.691(8)$, $1.15(7)$ and $1.5(2)$ \cite{Borinsky:2021jdb}. In $d=3$, our value also agrees with the fuzzy-sphere result $\Delta_{\phi^3}=4.613(6)$, i.e., $\omega=1.613(6)$ \cite{ArguelloCruz:2025zuq}. In $d=5$, the fuzzy four-sphere gives $\Delta_{\phi^3}=5.75(2)$, i.e., $\omega=0.75(2)$ \cite{Qian:2026fbx}, somewhat above both our result and the $\eps$ expansion. The older estimates of Ref.~\cite{LaiFisher1995}, $1.72(8)$, $1.30(3)$ and $0.73(2)$, lie above ours in all three dimensions.

At $d_\sigma$ we find $\omega=1.72(8)$, including the derivative-expansion uncertainty. Below $d_\sigma$, $\omega$ continues to increase; we discuss this region in the next subsection.


\subsection{Below \texorpdfstring{$d_\sigma$}{d sigma}: two and one dimensions}
\label{sec:lowd}

We continue the running-point calculation from $d_\sigma$ to $d=1$ at $\ord{2}$, $\ord{4}$ and $\ord{4}_{W_1}$, with $\alpha=2$. The resulting curves are included in Figs.~\ref{fig:eta}, \ref{fig:sigma} and \ref{fig:omega}, and Table~\ref{tab:exponents} lists values at selected dimensions. We estimate the uncertainties as above $d_\sigma$, with three changes. The field-truncation error is the change from $N_Z=14$ to $16$ at $\ord{4}$, and from $12$ to $14$ at $\ord{2}$ and $\ord{4}_{W_1}$. At $\ord{4}$, the regulator error, half the spread over $\alpha\in[1,3]$, is evaluated at $N_Z=14$ for $d\le1.5$ and at $N_Z=12$ above. The regulator error is not available below $d\simeq1.04$, as at $N_Z=14$ a spurious complex pair of eigenvalues becomes relevant for $\alpha=3$ and our calculations become unreliable.

The field expansion converges more slowly as $d$ decreases. At $\ord{4}$, $\eta$ changes by less than $5\times10^{-3}$ from $N_Z=14$ to $16$ down to $d\simeq1.3$; at $\ord{2}$ and $\ord{4}_{W_1}$, by less than $5\times10^{-3}$ from $N_Z=12$ to $14$ down to $d\simeq1.6$. For $\omega$, the change from $N_Z=14$ to $16$ stays below $0.05$, and that from $12$ to $14$ below $0.15$, only down to $d=2$. Below these dimensions, Figs.~\ref{fig:eta}--\ref{fig:omega} show the $N_Z$ dependence as a hatched band, and we regard our results in this region as qualitative.

In $d=2$, we find $\eta=-0.795(20)\pm0.007$ at $\ord{4}$, in agreement with the exact value $-4/5$ \cite{Cardy:1985yy}. For the correction-to-scaling exponent, we obtain $\omega=1.90(5)\pm0.03$, compared with the exact value $\omega=2$. Since $\omega$ is not determined by $\eta$, this comparison is an independent test of the derivative expansion at $d=2$.

In $d=1$, the $\ord{4}$ result $\eta=-0.97$ is consistent with the exact $\eta=-1$, with a field-truncation uncertainty of $0.04$. Since the field expansion is not converged below $d\simeq1.3$, these values are qualitative. A quantitative result near $d=1$ would require higher orders of derivative expansion and higher field orders or a different representation of the field dependence.

\section{Summary and outlook}
\label{sec:discussion}

We have determined the critical exponents of the Yang--Lee edge singularity using the functional renormalization group at second and fourth order in the derivative expansion.
Two features distinguish our work from earlier FRG studies:
First, \emph{complete derivative expansions:}
We retain all operators at $\ord{2}$ and $\ord{4}$. Omitting two of the three four-derivative operators changes the fourth-order correction by up to 50\%, and the sign of the discrepancy depends on dimension.
Second, \emph{a convergent representation of the field dependence:}
For the complex fixed point considered here, expansions about $\phi=0$ fail to converge with our regulator. Expanding about imaginary field values resolves this problem. In particular, the running massless point used by AMS gives convergent expansions down to $d\simeq1.3$.

Our fourth-order values of $\eta$ agree, within the quoted uncertainties, with the six-loop $\eps$-expansion results in $d=5$, $4$ and $3$ \cite{Gracey:2025rnz}. The correction-to-scaling exponent, $\omega=0.701(8)$, $1.20(4)$ and $1.59(3)$ in $d=5$, $4$ and $3$, is not fixed by $\eta$ and therefore tests the truncation independently. It agrees with the six-loop constrained-Pad\'e estimates \cite{Gracey:2025rnz} and, in $d=3$, with the fuzzy-sphere result \cite{ArguelloCruz:2025zuq}; as for $\eta$, the step from $\ord{2}$ to $\ord{4}$ moves it toward these values. In $d=2$, they agree with the exact values of $\eta$, $\sigma$ and $\nu_c$, and the correction-to-scaling exponent $\omega=1.90(6)$ is close to the exact $\omega=2$. Towards $d=1$ the field expansion converges more slowly; there our results are only qualitatively consistent with the exact values.

The main remaining limitation is the uncertainty associated with the derivative expansion itself. At present, we estimate it from the difference between second and fourth order. We have derived the strict $\ord{6}$ flows for twelve of the thirteen coefficient functions. Deriving the remaining flow and solving the resulting equations numerically will require a more efficient evaluation of the large expressions. A sixth-order calculation would allow us to apply the error-estimation methods developed for the Ising model \cite{Balog:2019rrg}.

Our determination of $d_\sigma$ agrees with the six-loop $\eps$-expansion estimate $2.649(25)$ \cite{Gracey:2025rnz,GraceyPrivate}.

\section{Addendum}
\label{sec:addendum}

After completing this work, we found the Mathematica code accompanying the arXiv submission of Ref.~\cite{An:2016lni}. Comparing its implementation with ours allowed us to identify the source of the reported eigenvalue-relation violations.

The code evaluates the flows of the Taylor coefficients $U^{(n)}(i\chi)$ and $z^{(n)}(i\chi)$ at fixed field, omitting the contribution $F^{(n+1)}(i\chi)\,\partial_t(i\chi)$ from the motion of the expansion point. Since this point runs with $k$, the chain rule requires this contribution in the flow of every Taylor coefficient. In Ref.~\cite{An:2016lni}, however, it is retained only in the equation for $\eta$.

The omission leaves the fixed-point solution \emph{unchanged}, because $\partial_t\chi=0$ at the fixed point. It does \emph{change} the linearized flow: the stability matrix no longer has $\Dphi$ as an exact eigenvalue. Removing the same contribution from our equations reproduces the large violations reported by AMS. We obtain 15\% in $d=5$ and 74\% in $d=4$, with the violation increasing to 200\% in $d=3.5$.

\begin{acknowledgments}
We thank J.~A.~Gracey for correspondence and for the six-loop estimate of $d_\sigma$. We also thank Piero Beretta and Carlos Sánchez-Villalobos for discussions on FRG and YLE. V.S. is indebted to ITP Heidelberg for hospitality and for creating a stimulating environment that led to this work. K.K. acknowledges funding by the GSI Helmholtzzentrum für Schwerionenforschung and by HGS-HIRe for FAIR. This research is supported in part by the ExtreMe Matter Institute EMMI at the GSI Helmholtzzentrum für Schwerionenforschung, Darmstadt, Germany. This work is funded by the Deutsche Forschungsgemeinschaft (DFG, German Research Foundation) under Germany's Excellence Strategy EXC 2181/1 - 390900948 (the Heidelberg STRUCTURES Excellence Cluster) and the Collaborative Research Centre SFB 1225 - 273811115 (ISOQUANT). The derivation, numerical code and data that support this work are available from the authors upon reasonable request.
\end{acknowledgments}

\appendix

\section{Outline for the derivation of the derivative-expansion flows}
\label{app:derivation}

This appendix describes the algorithm implemented in our Mathematica code. Given a local operator basis for a single scalar field, it generates the functional flow equations for all coefficient functions without imposing an internal symmetry. The dimension $d$ remains a symbolic parameter throughout the calculation.

\subsection{Vertices of the ansatz}

We write each operator in Eq.~\eqref{eq:ansatz} as $F(\phi)\,\mathcal{O}[\partial\phi,\partial^2\phi,\dots]$. The derivative structure $\mathcal{O}$ consists of $n_{\mathcal O}$ slots, each representing a field with a specified set of derivatives, contracted in a prescribed pattern. We expand around a uniform background, \[ \phi(x)=\phi+\sum_i\epsilon_i e^{ip_i\cdot x}, \] and extract the coefficient of $\epsilon_1\cdots\epsilon_n$ to obtain the $n$-point vertex:
\begin{multline}
  \Gamma^{(n)}(p_1,\dots,p_n;\phi) =\sum_{\mathcal O}F^{(n-n_{\mathcal O})}(\phi)\\ \times\sum_{\sigma} f_{\mathcal O}\big( p_{\sigma(1)},\dots,p_{\sigma(n_{\mathcal O})} \big).
\end{multline}
Here $\sigma$ runs over injective maps from the operator slots to the external legs. The function $f_{\mathcal O}$ contains the momentum factors $ip$ generated by the derivatives, together with the operator's explicit normalization. For example, \[ f_Z(k_1,k_2)=-\tfrac12 k_1\!\cdot\!k_2, \] \[ f_{W_2}(k_1,k_2,k_3) =\tfrac12(k_1\!\cdot\!k_2)k_3^2, \] and \[ f_{W_3}(k_1,\dots,k_4) =\tfrac12(k_1\!\cdot\!k_2)(k_3\!\cdot\!k_4). \] All momenta are incoming, with $\sum_i p_i=0$.

\subsection{Vertex flows}

We expand $\mathrm{Tr}[\partial_tR_k(\Gamma^{(2)}_0+V)^{-1}]$ in the fluctuation-dependent part $V$ of the two-point function. Extracting the multilinear coefficient gives
\begin{multline}
  \partial_t\Gamma^{(n)}(p_1,\dots,p_n) =\frac12\sum_{(B_1,\dots,B_k)}(-1)^k \int_q \partial_tR_k(q)\,G(q)^2\\ \times\prod_{j=1}^{k} \Gamma^{(2+|B_j|)}\!\big( q+Q_{j-1},-q-Q_j,\{p_i\}_{i\in B_j} \big)\\
  \times\prod_{j=1}^{k-1}G(q+Q_j),
  \label{eq:vertexflow}
\end{multline}
where $(B_1,\dots,B_k)$ runs over all ordered set
partitions of $\{1,\dots,n\}$ and
\[
  Q_j=\sum_{l\le j}\sum_{i\in B_l}p_i, \qquad Q_0=0.
\]
The insertion of $\partial_tR_k$ singles out a position on the loop, so the block order must be retained. Equation~\eqref{eq:vertexflow} reproduces the usual diagrammatic expressions. For $n=2$, for example, it reduces to \[ \partial_t\Gamma^{(2)}(p) =\int_q\partial_tR\,G^2 \Big[ \Gamma^{(3)}G(q+p)\Gamma^{(3)} -\tfrac12\Gamma^{(4)} \Big]. \]

\subsection{Momentum expansion and angular average}

We introduce a counting parameter $\lambda$ by rescaling the external momenta as $p_i\to\lambda p_i$. The shifted propagators are expanded according to \[ G\big((q+Q)^2\big) =\sum_m\frac{G^{(m)}(q^2)}{m!} \big(2q\!\cdot\!Q+Q^2\big)^m. \] Every product is retained through order $\lambda^s$, where $s$ is the derivative order of the truncation. We then perform the angular averages using
\begin{equation}
  \Big\langle\prod_{a=1}^{2m}(q\!\cdot\!k_a)\Big\rangle
  =
  \frac{
    (q^2)^m
    \sum_{\rm pairings}\prod(k_a\!\cdot\!k_b)
  }{
    d(d+2)\cdots(d+2m-2)
  }.
\end{equation}
This reduces the loop expressions to radial integrands while keeping their dependence on $d$ explicit.

\subsection{Projection}

After eliminating $p_n$ by momentum conservation, we expand both sides of Eq.~\eqref{eq:vertexflow} in monomials of the independent invariants $p_i\cdot p_j$. Matching their coefficients gives a linear system, \[ A\,x=b, \] where $x$ contains the unknown flows $\partial_tF^{(j)}$ entering $\Gamma^{(n)}$, and $b$ contains the one-dimensional loop integrands.

For $n\ge3$, the system is overdetermined. We solve it using the Moore--Penrose left inverse and verify that every component of the residual $Ax-b$ vanishes identically at the integrand level. This checks that the momentum-expanded loop expression can be represented as the vertex of a local functional through $\mathcal{O}(\partial^s)$.

We also check consistency between different vertex orders. For example, the flows of $Z'$ and $W_1'$ extracted from $\Gamma^{(3)}$ must equal the field derivatives of the corresponding flows from $\Gamma^{(2)}$. Likewise, the flows of $Z''$, $W_1''$ and $W_2'$ extracted from $\Gamma^{(4)}$ must agree with the appropriate derivatives of the lower-vertex flows. All these identities hold at any $y$.

\section{Convergence in the field expansion}
\label{app:fieldconv}

We truncate the Taylor expansions~\eqref{eq:taylor} at $N_U=N_Z+3$, $N_{W_1}=N_{W_3}=N_Z-2$ and $N_{W_2}=N_Z-1$. These relative orders follow from the coupling between coefficient functions in the flow equations. For example, the flow of $U$ at order $n$ involves $Z$ at order $n-2$ through the term $Zq^2$ in the propagator. With these offsets, a single parameter, $N_Z$, controls the field truncation.

\begin{figure}[t]
  \centering
  \includegraphics[width=\columnwidth]{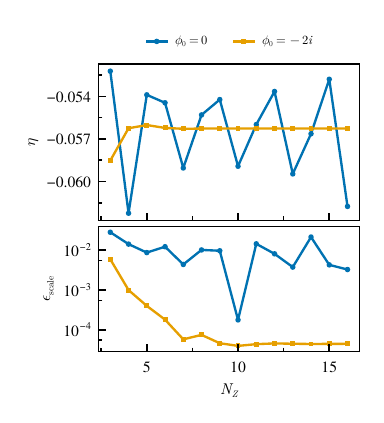}
  \caption{Convergence of $\eta$ with the field order at $d=5.5$, $\ord{2}$ and $\alpha=1$, for expansions about $\phi=0$ and $\phi=-2i$. The upper panel shows $\eta$; the lower panel shows the violation $\eps_{\rm scale}$ of the exact relation $\theta_t=\Dphi$. The period-three oscillation for the expansion about $\phi=0$ arises from complex singularities of the fixed-point functions.}
  \label{fig:fieldconv}
\end{figure}

Figure~\ref{fig:fieldconv} illustrates the effect of the expansion point. At $d=5.5$, the expansion about $\phi=0$ produces a non-decaying period-three oscillation as $N_Z$ increases. Moving the expansion point to $\phi=-2i$ instead gives rapid convergence, reaching seven-digit stability. A similar convergence pattern was found in Ref.~\cite{Pawlowski:2014zaa} in a different context.

\begin{figure*}[t]
  \centering
  \includegraphics[width=0.85\textwidth]{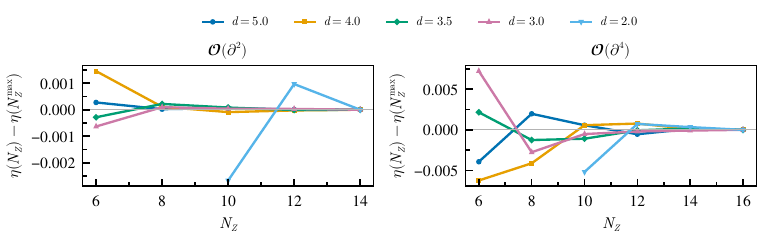}
  \caption{Convergence of $\eta$ with the field order $N_Z$ for the running massless point at $\alpha=2$. The left panel shows $\ord{2}$ and the right panel $\ord{4}$. Each curve gives the difference between $\eta(N_Z)$ and its value at the highest field order computed in that dimension. The reference order is $N_Z=14 (16)$ at $\ord{2}$ ($\ord{4}$). }
  \label{fig:nzconv}
\end{figure*}

For the running massless point, Eq.~\eqref{eq:massless}, convergence at $\alpha=2$ is shown in Fig.~\ref{fig:nzconv}. At $\ord{2}$, $\eta$ changes by less than $3\times10^{-4}$ beyond $N_Z=8$ in every dimension shown above $d_\sigma$. We use $N_Z=10$ for these calculations above $d_\sigma$.

At $\ord{4}$, convergence is slower. Near $d=3$, increasing $N_Z$ from 8 to 10 changes $\eta$ by $2\times10^{-3}$, while the change from 10 to 12 is $3\times10^{-4}$. Convergence deteriorates closer to six dimensions, where the running massless point approaches the origin. At $d=5$, the changes from $N_Z=10$ to 12 and from 12 to 14 are about $10^{-3}$ and $7\times10^{-4}$, respectively.

We therefore use fixed imaginary expansion points at $\ord{4}$ wherever they give convergent results, namely for $d\ge3.2$. For these expansions, the change from $N_Z=8$ to 10 is below $2\times10^{-4}$. The running-point results remain within $10^{-3}$ of the fixed-expansion-point results; at $N_Z=14$, the differences are $3\times10^{-4}$ in $d=5$ and $8\times10^{-4}$ in $d=4$.

Below $d=3.2$, we use the running massless point with $N_Z=12$ at $\ord{4}$. We estimate the field-truncation uncertainty from the change between $N_Z=10$ and 12. This change is at most $1.1\times10^{-3}$ over the dimensions considered in this range, and is $3\times10^{-4}$ at $d=3$.

We have also computed the running-point solution at $\ord{4}$ with $N_Z=14$ and $16$ over the whole range $1\le d\le 6$. Above $d_\sigma$, the two orders differ by $\le 3\times10^{-4}$ in $\eta$ and $\le 7\times10^{-3}$ in $\omega$.

Below $d_\sigma$, the field expansion converges more slowly as $d$ decreases and requires larger $N_Z$. At $\ord{2}$, the change from $N_Z=12$ to $14$ is $10^{-3}$ in $d=2$. In $d=2$, the $\ord{4}$ values of $\eta$ at $N_Z=10$, $12$, $14$ and $16$ are $-0.7997$, $-0.7938$, $-0.7942$ and $-0.7945$ (Fig.~\ref{fig:nzconv}); in $d=1.5$, $N_Z=14$ and $16$ give $-0.8915$ and $-0.8926$, while in $d=1$ they differ by $0.04$.

\section{Exact translation eigenvalue with the running massless point}
\label{app:translation}

Here we demonstrate that with the running massless point, Eq.~\eqref{eq:massless}, the stability matrix has the eigenvalues $\theta_t=\Dphi$ and $\theta_h=d-\Dphi$ \emph{exactly}, for any order of the field expansion. We give the explicit derivation in the LPA, where every step can be written out explicitly, and then explain why it carries over to $\ord{2}$ and $\ord{4}$. Eigenvalues $\theta$ are defined as the eigenvalues of $-M$, where $M_{ij}=\pd(\pd_tg_i)/\pd g_j$ is the stability matrix of the couplings $g_i$, so that relevant directions have $\theta>0$.

In the LPA, Eq.~\eqref{eq:flow} for the potential reads
\begin{equation}
  \pd_tU=-d\,U+\Dphi\,\phi\,U'+\ell(U''),
  \label{eq:lpa}
\end{equation}
where $\ell(m)=\int_0^\infty\dd y\,y^{d/2-1}I_U(y;m)$ is the loop integral evaluated with the mass $m$. Differentiating $n$ times with respect to $\phi$ gives
\begin{equation}
  \pd_tU^{(n)}=(-d+n\Dphi)\,U^{(n)}+\Dphi\,\phi\,U^{(n+1)}+\pd_\phi^n\,\ell(U'').
  \label{eq:lpan}
\end{equation}
The field enters explicitly only through the dilatation term $\Dphi\phi U^{(n+1)}$, while the loop term is a simple polynomial in $U^{(3)},\dots,U^{(n+2)}$.

The FRG couplings are the Taylor coefficients at the massless point, we denote them by $u_n\equiv U^{(n)}(i\chi)$. Because the expansion point moves with $k$, the flow of $u_n$ is the total derivative
\begin{equation}
  \pd_tu_n=\big[\pd_tU^{(n)}\big](i\chi)+u_{n+1}\,i\,\pd_t\chi.
  \label{eq:totalder}
\end{equation}
The second term describes the drift of the expansion point. Inserting Eq.~\eqref{eq:lpan}, the explicit $\chi$ of the dilatation term and the drift term combine into
\begin{equation}
  \pd_tu_n=(-d+n\Dphi)\,u_n+L_n+u_{n+1}\,\psi,
  \label{eq:un}
\end{equation}
with $\psi\equiv i\,(\pd_t\chi+\Dphi\,\chi)$ and $L_n\equiv\pd_\phi^n\ell(U'')|_{\phi=i\chi}$. Because $U''(i\chi)=0$, the loop derivatives are taken at zero mass, and $\ell^{(j)}\equiv\ell^{(j)}(0)$ are merely numbers that depend only on $d$, $\eta$ and the regulator, e.g.
\begin{align}
  L_0&=\ell,\qquad L_1=\ell^{(1)}u_3,\qquad L_2=\ell^{(1)}u_4+\ell^{(2)}u_3^2,\nonumber\\
  L_3&=\ell^{(1)}u_5+3\ell^{(2)}u_3u_4+\ell^{(3)}u_3^3. \label{eq:Ln}
\end{align}
Note that $L_n$ do not depend on $\chi$.

The condition $u_2=0$ holds for all $k$. Equation~\eqref{eq:un} for $n=2$ then fixes $\psi$:
\begin{equation}
  0=L_2+u_3\,\psi\quad\Longrightarrow\quad\psi=-\frac{L_2}{u_3}.
  \label{eq:psi}
\end{equation}
With this, the flow equations become
\begin{align}
  \pd_tu_n&=(-d+n\Dphi)\,u_n+L_n-u_{n+1}\frac{L_2}{u_3},\qquad n\ne2,
  \label{eq:unfinal}\\
  \pd_t\chi&=-\Dphi\,\chi+i\,\frac{L_2}{u_3}.
  \label{eq:chiflow}
\end{align}
Here one has to account for the field truncation $u_n=0$ for $n>N_U$. The right-hand side of Eq.~\eqref{eq:unfinal} then depends on $u_3,\dots,u_{N_U}$ (and on $u_0,u_1$ for $n=0,1$), but, most importantly, \emph{not} on $\chi$. The position $\chi$ appears only in Eq.~\eqref{eq:chiflow}; it enters there linearly and with exact coefficient $-\Dphi$!

\paragraph*{Explicit example: $N_U=4$.} With $u_5=u_6=0$, the complete set of flow equations is
\begin{align}
  \pd_tu_4&=(-d+4\Dphi)\,u_4+3\ell^{(2)}u_4^2+6\ell^{(3)}u_3^2u_4+\ell^{(4)}u_3^4,\nonumber\\
  \pd_tu_3&=(-d+3\Dphi)\,u_3+2\ell^{(2)}u_3u_4+\ell^{(3)}u_3^3-\ell^{(1)}\frac{u_4^2}{u_3},\nonumber\\
  \pd_tu_1&=(-d+\Dphi)\,u_1+\ell^{(1)}u_3,\nonumber\\
  \pd_tu_0&=-d\,u_0+\ell+u_1\psi,\nonumber\\
  \pd_t\chi&=-\Dphi\,\chi+i\,\frac{\ell^{(1)}u_4+\ell^{(2)}u_3^2}{u_3}.
  \label{eq:lpa4}
\end{align}
In the $u_1$ equation the drift term is absent because $u_2=0$. With the couplings ordered as $(u_4,u_3,u_1,u_0,\chi)$, the stability matrix is lower triangular in blocks,
\begin{equation}
  M=
  \begin{pmatrix}
    K_{44}&K_{43}&0&0&0\\
    K_{34}&K_{33}&0&0&0\\
    0&\ell^{(1)}&-(d-\Dphi)&0&0\\
    *&*&\psi&-d&0\\
    *&*&0&0&-\Dphi
  \end{pmatrix},
  \label{eq:Mlpa4}
\end{equation}
where $K$ is the $2\times2$ matrix of derivatives of the first two equations in Eq.~\eqref{eq:lpa4} and the entries marked $*$ do not affect the eigenvalues. The eigenvalues are therefore those of $-K$, which describe the irrelevant directions, together with the obvious
\begin{equation}
  \theta_h=d-\Dphi,\qquad \theta=d,\qquad \theta_t=\Dphi \,.
\end{equation}
The eigenvalue $d$ belongs to the field-independent part of the potential. The column of $M$ belonging to $\chi$ vanishes except for its diagonal entry, so the unit vector $e_\chi$ is an exact eigenvector. It describes the infinitesimal translation of the fixed point: $U(\phi)\to U(\phi-i\,\delta\chi)$ moves the massless point to $i(\chi+\delta\chi)$ and leaves every $u_n$ unchanged.

The same block structure holds for every $N_U$: the equations for $u_3,\dots,u_{N_U}$ form a closed subsystem, $u_1$ feeds only into $u_0$, and $\chi$ feeds into nothing. Hence $\theta_t=\Dphi$ and $\theta_h=d-\Dphi$ for \emph{every} truncation order.

\paragraph*{Comparison with a fixed expansion point.} For a fixed point $i\chi_0$,
there is no drift term and Eq.~\eqref{eq:un} is replaced by
\begin{equation}
  \pd_tu_n=(-d+n\Dphi)\,u_n+i\Dphi\chi_0\,u_{n+1}+\pd_\phi^n\ell(U'')|_{i\chi_0},
\end{equation}
with $u_2\ne0$. The translation corresponds to $\delta u_n=u_{n+1}$. Differentiating the fixed-point equation with respect to $\phi$ shows that this is an eigenvector with eigenvalue $-\Dphi$ of the \emph{untruncated} system. In a truncated system the components $\delta u_{N_U}=u_{N_U+1}=0$. The relation $\theta_t=\Dphi$ is then violated by an amount that decreases only as the Taylor series converges; this leads to a non-zero $\eps_{\rm scale}$ of Fig.~\ref{fig:diag}.

\paragraph*{Higher orders of the derivative expansion.}
At $\ord{2}$ and $\ord{4}$ each coefficient function $F$ obeys Eq.~\eqref{eq:flow}. Its loop term depends on the field only through the functions and their derivatives, and the dilatation term is again $\Dphi\phi F'$. For the Taylor coefficients $f_n=F^{(n)}(i\chi)$ the dilatation and drift terms therefore combine into $f_{n+1}\psi$, as in Eq.~\eqref{eq:un}. The two conditions defining the expansion point, $U''(i\chi)=0$ and $z(i\chi)=1$, give two equations, $0=L_2+u_3\psi$ and $0=\eta+z_1\psi+L_z$, where $z_1=z'(i\chi)$ and $L_z$ is the loop term of $z$. All coefficients in these equations are independent of $\chi$, and they determine $\psi$ and $\eta$. All flows of the Taylor coefficients are therefore independent of $\chi$, while $\pd_t\chi=-\Dphi\chi-i\psi$. The block structure of Eq.~\eqref{eq:Mlpa4} and the exact eigenvalues $\theta_t=\Dphi$ and $\theta_h=d-\Dphi$ follow in the same way as the LPA.

The anomalous dimension, and with it $\Dphi=(d-2+\eta)/2$, is a function of the Taylor coefficients, and we include this dependence when computing $M$. It contributes to the rows of $M$, including the row of $\chi$ through the term $-\Dphi\chi$, but not to the column belonging to $\chi$, because $\eta$ does not depend on $\chi$. The exact eigenvalues are therefore obtained whether or not the dependence of $\eta$ on the couplings is included. The other eigenvalues do depend on this choice. Including the dependence is required to reproduce the one-loop result $\omega=\eps+O(\eps^2)$; with $\eta$ held fixed at its fixed-point value, one finds instead $\omega\simeq4\eps/3$ near six dimensions.

\section{Full field dependence for the Ising test}
\label{app:collocation}

For the Ising benchmark in Sec.~\ref{sec:validation}, we resolve the full field dependence using Chebyshev collocation. We compactify the real field axis through \[ \xi=\frac{\phi^2}{\phi^2+\ell^2}\in[0,1) \] and factor out the known large-field behavior of each coefficient function, \[ F(\phi) =\left(1+\frac{\phi^2}{\ell^2}\right)^{p_F/2} \tilde F(\xi), \] where $p_F$ is its large-field power. We expand $\tilde F$ in Chebyshev polynomials, with $\mathbb{Z}_2$ symmetry incorporated into the basis, and impose the fixed-point equations at the Chebyshev nodes. The critical exponents are then obtained from the eigenvalues of the discretized stability matrix.

This method also works for the Yang--Lee fixed point close to six dimensions. At $d=5.95$, it reproduces the Taylor-expansion result for $\eta$ to seven digits. Below $d\approx5$, however, the collocation problem becomes ill-conditioned.

For the Ising model, the large-field behavior along the real axis selects the physical solution. For the complex Yang--Lee solution, this condition is no longer sufficient, and the discretized equations admit nearly degenerate spurious solutions. Selecting the physical branch requires control of analyticity in the complex field plane. A Taylor expansion incorporates local analyticity directly, making it better suited to the present calculation. We therefore use Taylor expansions for all Yang--Lee results.

\section{Uncertainty estimates}
\label{sec:errors}

We have four major sources of uncertainty.

\paragraph*{Field truncation.}
We estimate the uncertainty from the finite Taylor expansion by the change in $\eta$ when $N_Z$ is increased by two. This change is below $5\times10^{-4}$ at $\ord{2}$ and below $3\times10^{-3}$ at $\ord{4}$ (Appendix~\ref{app:fieldconv}) for $d\ge d_\sigma$. Below $d_\sigma$, we use the change from $N_Z=14$ to $16$ at $\ord{4}$ and from $12$ to $14$ at $\ord{2}$ (Sec.~\ref{sec:lowd}).

\paragraph*{Regulator dependence.}
We take half the spread of $\eta$ over $\alpha\in[1,3]$ as the regulator uncertainty. When a PMS point is available (Sec.~\ref{sec:regulator}), we instead use the shift from the $\alpha=2$ result to the PMS value if it is larger. This applies only at $\ord{4}$ in $d=4$, where the shift is $1.7\times10^{-3}$.

\paragraph*{Normalization point.}
At finite order in the derivative expansion, $\eta$ depends weakly on the field value at which $z$ is normalized, since the regulator $R_k\propto Z_0$ is not invariant under field rescalings. We estimate this uncertainty from the difference between the two field representations described in Sec.~\ref{sec:numerics}. It is below $3\times10^{-4}$ for $d\ge4$ and reaches $3\times10^{-3}$ near $d=3.4$. This uncertainty is most likely correlated with the regulator dependence. Below $d_\sigma$, only the running massless point converges, and this estimate is not available.

\paragraph*{Derivative expansion.}
The finite order of the derivative expansion is the dominant source of uncertainty. For the Ising model, the change in $\eta$ from $\ord{4}$ to $\ord{6}$ is several times smaller than that from $\ord{2}$ to $\ord{4}$ \cite{Balog:2019rrg}. Assuming a similar reduction here, we assign the fourth-order result an uncertainty of \[ \delta_{\rm DE}\eta =\frac{1}{3}\big| \eta_{\ord{4}}-\eta_{\ord{2}} \big|. \]

\bibliography{refs}

\end{document}